# AUTOMATIC DETECTION OF DEPRESSION FROM STRATIFIED SAMPLES OF AUDIO DATA

**Pongpak Manoret [1,¶], Punnatorn Chotipurk [1,¶], Sompoom Sunpaweravong [1,¶], Chanati Jantrachotechatchawan[2], Kobchai Duangrattanalert [3*]**

[1] Triam Udom Suksa School, Thailand;
  pongpak282546@gmail.com (P.M.), pun.chotipurk@gmail.com (P.C.);
  sompoomsunpaweravong@gmail.com (S.S.)

[2] Research Division, Faculty of Medicine Siriraj Hospital, Mahidol University, Thailand;
  chanati.jan@mahidol.ac.th (C.J.)

[3] Chulalongkorn University Technology Center (UTC), Chulalongkorn University Thailand;
  kobchai.d@chula.ac.th (K.D.)

¶ These authors contributed equally to this work
* Correspondence

**Abstract:** Depression is a common mental disorder which has been affecting millions of people around the world and becoming more severe with the arrival of COVID-19. Nevertheless proper diagnosis is not accessible in many regions due to a severe shortage of psychiatrists. This scarcity is worsened in low-income countries which have a psychiatrist to population ratio 210 times lower than that of countries with better economies. This study aimed to explore applications of deep learning in diagnosing depression from voice samples. We collected data from the DAIC-WOZ database which contained 189 vocal recordings from 154 individuals. Voice samples from a patient with a PHQ-8 score equal or higher than 10 were deemed as depressed and those with a PHQ-8 score lower than 10 were considered healthy. We applied mel-spectrogram to extract relevant features from the audio. Three types of encoders were tested i.e. 1D CNN, 1D CNN-LSTM, and 1D CNN-GRU. After tuning hyperparameters systematically, we found that 1D CNN-GRU encoder with a kernel size of 5 and 15 seconds of recording data appeared to have the best performance with F1 score of 0.75, precision of 0.64, and recall of 0.92.

# 1. INTRODUCTION

Depression is a common psychiatric disorder which has been affecting millions around the world for many decades. The severity of this crisis worsened with the arrival of COVID-19 outbreaks around the world with the global death toll rising from 264 million in 2018 to 280 million 2021 [1]. This high prevalence of depression accounts for 3.76% of the global population and 5.02% of the adult population. Depression diagnosis and treatment has been inaccessible in many parts of the world due to financial costs, privacy concerns and a severe shortage of psychiatrists. This scarcity is worsened in low-income countries which have a psychiatrist to population ratio 210 times lower than that of countries with better economies [2]. As a result, up to two-thirds of all depression cases are left undiagnosed and thus untreated [3]. Untreated depression can ruin quality of life by causing loss of sleep, concentration, and happiness. In the worst cases, depression can lead to suicide, which claims 700,000 lives every year and is the fourth leading cause of death in 15-29 year olds [4]. Furthermore, depression is also the leading cause of disability worldwide [5]. From a socioeconomic standpoint, depressed individuals lose up to 4 more hours of productive work per week compared to their healthy counterparts [6]. This can have devastating impacts on both the individual's livelihood and the overall economic output which is estimated to cost $210.5 billion per year globally [7]. Therefore, it is crucial to make depression diagnosis more accessible to mitigate these consequences on society as well as individuals.

Current depression diagnostic methods range from clinical diagnosis using interviews with a psychiatrist to questionnaires such as PHQ-2 [8], PHQ-8 [9] and PHQ-9 [10]. While being diagnosed by a psychiatrist can detect a wider range of expressions and symptoms of the disease, it lacks scalability and is constrained by the shortage of psychiatrists to conduct the interview. As reported by the World Health Organization, there are only 0.05 psychiatrists per 100,000 people in low-income countries [2]. Although this proportion is larger in high-income countries, it is still low with a ratio of 10.5 psychiatrist per 100,000 people. Patient health questionnaires are less practical in diagnosing depression due to their low sensitivity and specificity compared to clinical diagnosis methods. Taking the PHQ-8 questionnaire as an example, it has a sensitivity of 77% and a specificity of 62%, which is significantly lower than structured clinical interviews which have a sensitivity of 95% and a specificity of 84% [11][12]. This presented an opportunity to apply automatic speech recognition technologies to diagnose depression. Not only does this method reduce reliance on human resources, but it is also more scalable and affordable than in-person screening.

Over the last decade, numerous technologies have been developed to detect various kinds of mental disorders including anxiety [16], PTSD [17] and Alzheimer's disease [18]. In addition to this, many related works have experimented with the use of artificial intelligence to diagnose depression through voice [19][20][21][22]. However, there is still room for improvement as various model architectures have not been explored for this particular task.

This paper proposes a new method of using deep learning to conduct speech-based depression screening. We build a classification model on DAIC-WOZ dataset [18] and perform rigorous experimentations to determine potentials of our proposed strategy.

After searching rigorously for an optimal model configuration, we found that 1D CNN-GRU encoder with a kernel size of 5 and 15 seconds of recording data appeared to have the best performance with F1 score of 75.00 ± 0.82%, sensitivity of 91.67 ± 5.26%, and precision of 64 ± 2.36%.

## 2. RELATED WORK

### 2.1. Depression Screening and Diagnosis

Depression is defined by the American Psychiatric Association as a common mental disorder that causes sadness and loss of interest in activities which were once enjoyed by the individual [4][62]. It is distinct from sadness experienced regularly as part of life by its worryingly long and severe emotional disturbance. Depression displays symptoms that last over two weeks and reportedly have strong effects on both functionalities and behaviors of the patient.

Depression patients also suffer from the thoughts of being worthless which can lead to self-inflicted harm or even suicide. Many risk factors play a role in causing depression such as environment, socioeconomic factors and adverse life events (unemployment, traumatic events etc.) [4]. It is estimated that 1 in 15 adults are affected by depression annually and 1 in 6 people will encounter depression at some points in their life. However, there are multiple effective methods of treating depression ranging from antidepressant medication to interpersonal psychotherapy [4].

Conventionally, mental disorders including depression are diagnosed manually by psychiatrists. Current methods consist of interviews and questionnaires (such as PHQ-2 [8], PHQ-8 [9] and PHQ-9 [10]) designed to diagnose this type of disorder. Results from these surveys will be analyzed by psychiatrists who subsequently conduct an one-on-one interview with the patient. During the interview, psychiatrists search for markers related to depression in patients' speech e.g. emotional display, reasonings, and inconsistencies [25]. Manual depression diagnosis is proven to be effective, but there are not enough psychiatrists available, and in-person diagnosis is generally time-consuming [13]. Furthermore, the financial burden in major depression disorders is relatively high [11][12]. Fortunately, several medical technologies have been developed to tackle these problems. Currently, there has been an integration of Artificial Intelligence (AI) in various fields of medical analysis [26][27][28][29][30], but not many were practically used in psychological disorders detection. Skills developed by psychiatrists for instance speech pattern analysis can be learned and mastered by AI. Therefore, AI has become a promising alternative to manual depression detection.

The PHQ-8 is an eight-item patient health questionnaire commonly used to measure severity of depression [9]. Each question within the questionnaire asks how frequently respondents experience certain symptoms which are commonly found in depressed patients. The frequency of these symptoms directly correlates with the score given for each question. Those who don't experience the symptoms at all are scored 0 and those who experience the symptom on a near-daily basis are given a score of 3 (see Table 1).

Table 1. PHQ-8 scores and depression severity [9]

| PHQ-8 Score | Severity Level |
|---|---|
| 0-4 | Minimal depression |
| 5-9 | Mild depression |
| 10-14 | Moderate depression |
| 15-19 | Moderately severe depression |
| 20-24 | Severe depression |

Speech is an act of expressing ideas and emotion by vocalization [31]. It is also an indispensable component for communication between individuals inside human society. As for communication, another element called "language" has been used along with speech. Language is the way to express thought through a distinct set of symbols, dialects, or sounds (speech). Language understanding can be acquired by a comprehensive study of vocal patterns and alphabets. Humans are capable of identifying as well as expressing speech and languages; meanwhile, machines do not have the ability to do so.

For automatic speech recognition (ASR), the system will process vocal data (speech) into digital signals suitable for AI training and analysis [32]. Speakers have unique voice patterns due to the variation of personalities and body structure. Accordingly, ASR uses criteria such as speech size and speaking styles to classify voice samples into groups. Spectrograms and chroma feature techniques may potentially enhance the system in organizing voices [33]. Both techniques extract and present relevant features to the system, enabling the system to conduct more complex classification and evaluation [34][38][42][52]. The methods provided above allow AI to effectively perform speech and emotions recognition.

**2.2. Artificial Intelligence for Speech Recognition**

A range of research has been conducted in the field of speech classification and recognition. In 2013, Li Deng and colleagues [35] presented an overview of "New Types of Deep Neural Network Learning for Speech Recognition and Related Applications," which is a collection of studies related to sound technologies. Li Deng's and his team introduced the historical accounts of speech-related neural networks development. Prior to the publication of this paper, DNNs ( Deep Neural Networks) had been introduced to improve the GMMs (Gaussian Mixture Model) in the field of acoustic models. Although DNNs outperformed the GMMs, there were, at that time, few ways to train the DNNs effectively. The summary focused on various ways to improve deep neural networks. The process can be classified into five methods 1. Preprocessing of speech into spectrograms, 2. Enhancement of neural networks activation functions, 3. Improvement of ways to learn the DNNs parameters, 4. Expansion of optimization strategies, and 5. Development of machines' ability to leverage multiple languages simultaneously. Moreover, the research also concluded that DNNs can be extensively used in areas further than speech recognition, such as computer vision and natural language processing.

Over the past years, speech recognition has received a number of improvements. Haizhou Li et al. 2013 [36] described the nature of speech recognition DNNs from both phonological and computational perspectives. Ganesh K. Venayagamoorthy et al. [37] introduced spectrograms to demonstrate pattern distinction between two speech samples. The team also trained neural networks to recognize speech patterns as well as speakers. Both experiments revealed that despite advancements in speech recognition there was room for improvement especially in assessing the state of emotion.

Later research carried out by Ismail Shahin [38] is an example demonstrating the deep neural network performance to identify speech under different emotion patterns. From the paper, speaker recognition was divided into speaker identification (SI) and speaker verification (SV). SI is the process of determining which speech pattern groups match the unknown vocal input. Speaker verification (SV) is the process of accepting or rejecting an identity of the group which will be performed after the classification of sample voices is completed. Shahin aimed to implement these 2 subdivided speakers recognition processes into DNNs models. He used an acoustic database composed of sentences uttered in 6 distinct emotion groups: neutral, angry, sad, happy, disgust, and fear. The CHMM2s model, which is a probabilistic based system that can predict outcome variables from the internal sequences that can be observed in a sound, was employed as a classifier: the main model for training and evaluating the outcomes. After the operation, the model achieved a performance of 79.92% accuracy when analyzing the speaker's sentiment identification. Though the result obtained was not perfect, Shahin's research laid a foundation for future acoustic-based DNNs projects in psychological fields.

## 2.3. Artificial Intelligence for Psychological Diagnosis

Over the past few years, speech recognition artificial intelligence has been used as an alternative for manual mental disorders detection. Research conducted in 2018 [41] demonstrated deep learning networks' ability to detect mood disorders (unipolar depression and bipolar disorder). The system was trained using speech responses from participants who watched videos that evoke six distinct emotions. By embodying CNNs (Convolutional Neural Networks) and attention-based RNNs (Recurrent Neural Networks), the team was able to generate emotion profiles (EPs) from various vocal data. Next, the system will evaluate and highlight speech patterns from the EPs that are relevant to mood disorders. For the outcomes, the model had achieved the accuracy of 0.87 and 0.76 for emotions and mood disorders detection respectively.

Another approach to use DNNs in psychological disorders analysis was made in 2018. Tuka Alhanai et al. [42] utilized Long-Short Term Memory (LSTM) neural networks to differentiate depression patients from ordinary people during an emotional talking environment. It also examined the severity of depression through acoustic analysis. The procedure encompassed modeling of sequences from both patients' voices and questionnaires so that the LSTM could evaluate the global patterns of the patients' voice [39]. Based on the data extracted, three separate experiments were carried out with specific outlines: 1. Given/Not given the condition of questions asked (using Logistic Regression system) or 2. Using LSTM without knowing the types of questions. For the result, the LSTM-based model achieved a 0.77 F1 score (harmonic average of recall and precision) when evaluating depression in patients. Tuka Alhanai et al. also integrated facial expression into their diagnostic model to further enhance its accuracy.

Recently, Adrian Vasques et al. [43] proposed a new depression analytic system using CNNs. The project was carried out by using DAIC-WoZ dataset which contains vocal data for depression analysis. Next, the acoustic input was downsampled and transformed into log spectrograms in the form of Short-Time Fourier Transform (STFT). Subsequently, the preprocessed data entered a model-analysis stage which encompassed a 1-D CNNs classifier and ensemble techniques. Such a combination produced scores that were stabilized and made more reliable by 5-fold-cross-validation. Consequently, the model achieved 0.65 F1 score, 0.53 precision and 0.71 recall in depressed patients.

**2.4. Summary**

As previously stated, artificial intelligence has been continuously improved and implemented into medical domains over the years. The findings that we reviewed above had laid the foundation for the development of depression detection AI that our team wished to be a part of. They have demonstrated that depression can pragmatically be detected with speech recognition DNN. Nevertheless, there are always rooms for improvement or even potentially different approaches. Currently, people have limited access to proper diagnosis due to the insufficient number of professionals and unsuitable methods for detecting such conditions [13]. With this foundation of related works reviewed, we hope to improve upon these relevant studies and implement their findings to improve all aspects of our study.

# 3. MATERIAL AND METHODS

## 3.1 Data Source and Description

DAIC-WOZ is part of a larger dataset - DAIC (Distress Analysis Interview Corpus) which compiles hundreds of interview data designed to support the study of multiple psychological distress conditions such as depression, anxiety, and post-traumatic stress disorder [22]. This dataset is provided by the University of Southern California and contains 189 voice recording samples from 154 individual subjects [22]. The data were collected by conducting interviews with the participants through an animated virtual interviewer. The data collected included audio and video recordings of the participant's expressions which were paired with questionnaire responses and scores. Subjects were labeled with a single value per recording as healthy or depressed using PHQ-8 scores as the criteria. Those with PHQ-8 scores lower than 10 were labeled as healthy and those with scores over 10 were labeled as depressed. Of the 154 subjects, 44 were labeled as depressed and the remaining 110 were considered as healthy control subjects. The length of each voice recording varied between 7 and 33 minutes with the average length being 16 minutes. Each participant made between 42 and 386 responses throughout their interview. The length of each response varies between 1 and 21 minutes.

## 3.2 Data Preprocessing

We downsampled all the voice samples to 8,000 Hz. Subsequently, we have a hypothesis that patients' speech will be more emotionally engaged if the vocal records are long enough. Short-term voice recordings may have very limited information useful for detecting depression, so we decide to use long recordings instead. Based on the given dataset, the size of all samples varies between 2 and 7 seconds long. Regarding the hypothesis, we set the sample size threshold to be at least 3 seconds per record (slightly below average length of all samples); those in the dataset that do not meet such criteria will be eliminated before being input to the system.

Techniques for feature extraction such as Mel-Spectrograms have been used extensively in the field of speech recognition systems [44][45][46][47]. Mel spectrogram is a non-linear transformation of frequency based on the waveform of the audio samples [48]. Mel-spectrograms are calculated based on logarithmic frequency spacing and frequency amplitude. On the spectrogram, magnitude will indicate the amplitude of the extracted frequencies [49][50]. However, when portrayed on a Mel-Spectrogram, the values representing amplitudes of each frequency domain (features) at a given time interval are often too subtle to distinguish. In order to emphasize the features, we convert amplitude on Mel-Spectrograms to decibel scale (see Figure 1). After this stage, distribution in each Mel-Spectrogram row becomes more pronounced; hence, once we input them into the model, the system will be capable of identifying key features more easily [49].

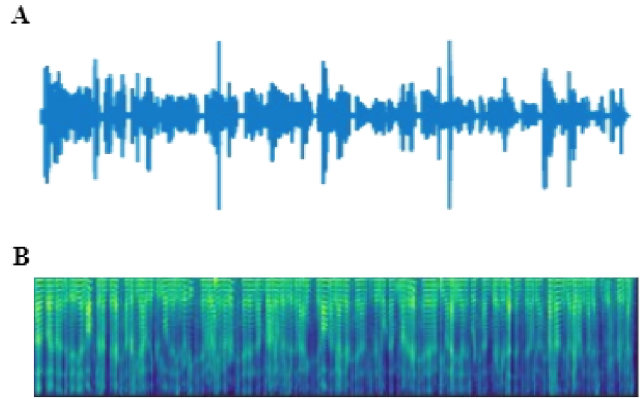

**Figure 1.** An example of audio data after being downsampled to 8,000 Hz in (A) waveform and in (B) Mel-Spectrogram.

Next, we split all the samples into equally sized chunks which are 1 second (16 units) long. During the process, we divide fragments to overlap each other by 15 units from the recording of every independent person in the dataset; the aforementioned method will expand the sample pool for the model.

In many speech recognition projects, the analysis focuses on the distribution of acoustic patterns over time [43][48][51][58]. However, we have another hypothesis that, apart from patterns across the time axis, there might be markers related to depression distributed across the frequency domain. Based on this concept, we propose that abnormal distribution of these markers may potentially be the sign of depression. Therefore, we decided to test the idea by transposing the Mel-Spectrogram samples from 16 x 128 (time x frequency) to 128 x 16 (frequency x time). In this manner, the model will be capable of searching for key features and their association in the frequency domain.

### 3.3. Model Architecture

For our hypothesis, we believe that people with depression might have biomarkers distributed sparsely across his/her vocal recording. Therefore, we expect that by providing random samples to the model, the chance of detecting the biomarkers will increase. As for our project, we employ a generator that randomly selects fragments from a recording of each person for analysis. Once the samples are selected, they will enter an encoder and further layers later on. Meanwhile, the Time Distributed Layer will simultaneously encode each fragment in a parallel manner into a stack of vectors (see Figure 2). Time Distributed Layer compiles all the feature vectors into a 2-dimensional array. (fragments were piled up into a stack that has length and depth). However, it might not be not appropriate to feed the resulting matrix directly into the Multilayer Perceptron (MLP). The reason is that these features appear to be equally important when it comes to detecting the disease. Every fragment has a vector that represents its frequency distribution and other acoustic properties.

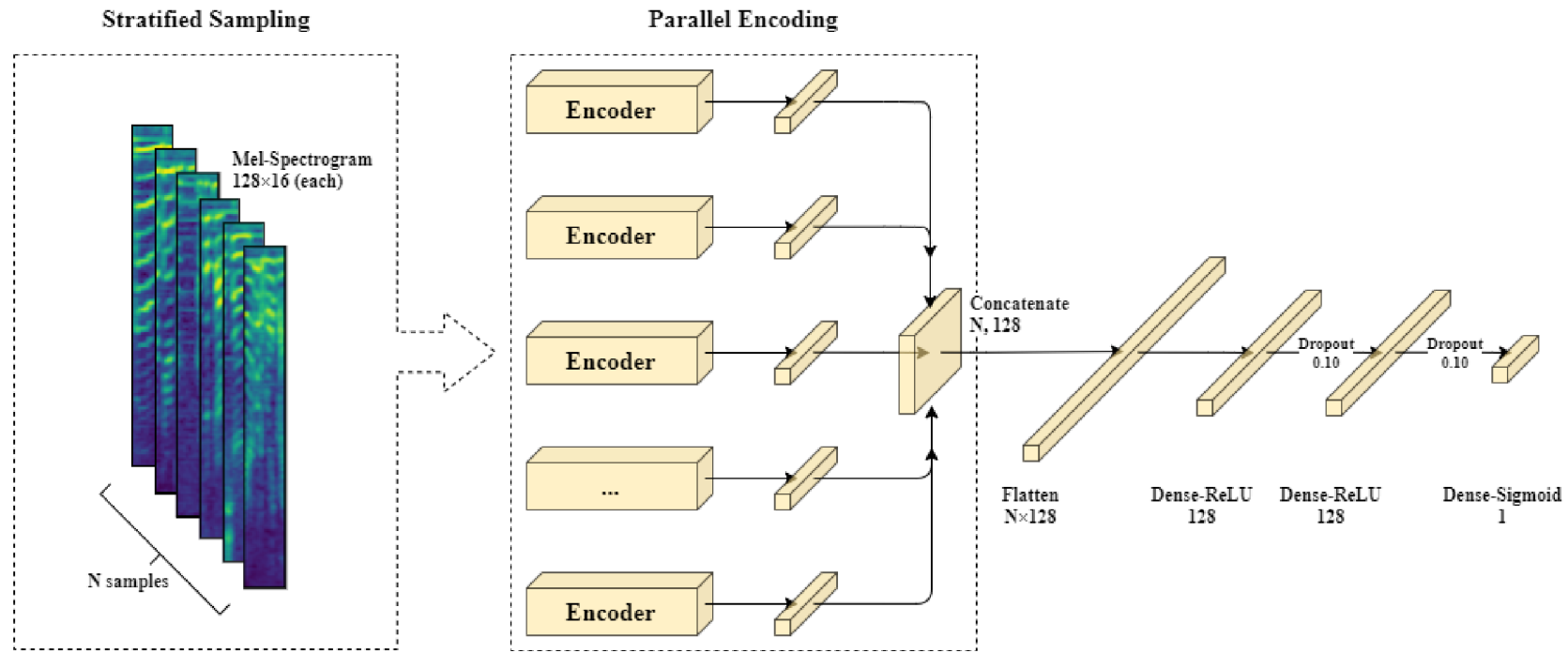

**Figure 2.** Architecture of the proposed model - First, N fragments of Mel-Spectrogram (Input) enter the model independently from each other while they are supervised by TimeDistributed Layer. N represents the number of vocal fragments selected randomly by the model (each piece has the size of 128 x 16). Here, these fragments are encoded simultaneously to generate corresponding feature vectors that are unique to each voice fragment. Next, these vectors will be concatenated into a stack of N, 128. After that, Flatten transforms the N, 128 array into an N x 128 feature vector which is then input to the dense layers. Dense layers which are set with 0.10 dropout rate summarize the feature vector (convert from N x 128 to 128). For the Sigmoid Activation Function, it evaluates a number (Output) between 0 (healthy) and 1 (depressed).

With this in mind, our aim in this stage is to vectorize our fragments (2D to 1D) without undermining their significance. So, we decide to not employ global pooling layers but instead use Flatten to prevent information loss.

Once we obtain a feature vector from Flatten, it will be input into a stack of dense layers. Here, the MLP will adjust the weight and summarize every value (number that represents the feature) on the vector. To prevent overfitting, we employ Dropout with 0.1 dropout rate as a regularizer. After that, in the final layer of the MLP, we apply the Sigmoid Activation Function to the layer for binary classification. The output of the dense layer will be calculated by the Sigmoid Activation Function and reported as a real number between 0 and 1. If the person is a depressed subject, the outcome will be a number close to 1. Meanwhile, if the subject is healthy, the reported number will be close to 0. However, our model does not use the threshold number of 0.50 to distinguish depressed subjects from healthy ones as we want to avoid complications coming from imbalance classes within the dataset. Therefore we decided to use Area under Curve (AUC) as our metric during training. In general, AUC can be derived from 2 distinct approaches - 1. Receiver Operating Characteristic (ROC) curve and 2. Precision-Recall (PR) curve [63][64]. In this project, we will calculate this score from the PR curve as we want to focus mainly on precision and recall of the model.

And finally, the model is compiled with PR-AUC as metric, AdabeliefOptimizer (learning rate: 1e-4; epsilon: 1e-7; rectify: True) as optimizer, and Sigmoid Focal Cross Entropy (alpha: 0.25; gamma: 2.00) as loss function.

Hyperparameters are internal variables that determine the performance and structure of the model. They cannot be learned in a model training process, so humans and additional algorithms are required to adjust the values of these parameters. In this project, we use grid search training: a hyperparameter tuning procedure that lists all the combinations of parameters, then automatically runs the model to search for the most yield outcome (i.e. highest AUC). Here, we will select our final model (a combination of parameters) from the one that shows the best performance.

Encoder has primary functions to extract and summarize features from the input data. In this project, we hypothesize that biomarkers will appear over the frequency axis but we do not know whether they are local patterns or global patterns. Hence, in order to handle this issue relating to such patterns, each encoder has 1D Convolutional Neural Network (1D CNN) with or without LSTM or GRU (see Figure 3). The selected Mel-Spectrogram samples will be fed into 1 out of 3 available encoders (Type of encoder will be discussed in detail later in section 3.4.). After that, the input will be vectorized by GlobalMaxPooling and normalized subsequently by a stack of Dense layers. As a result, feature vectors that represent acoustic data of those selected samples will be generated.

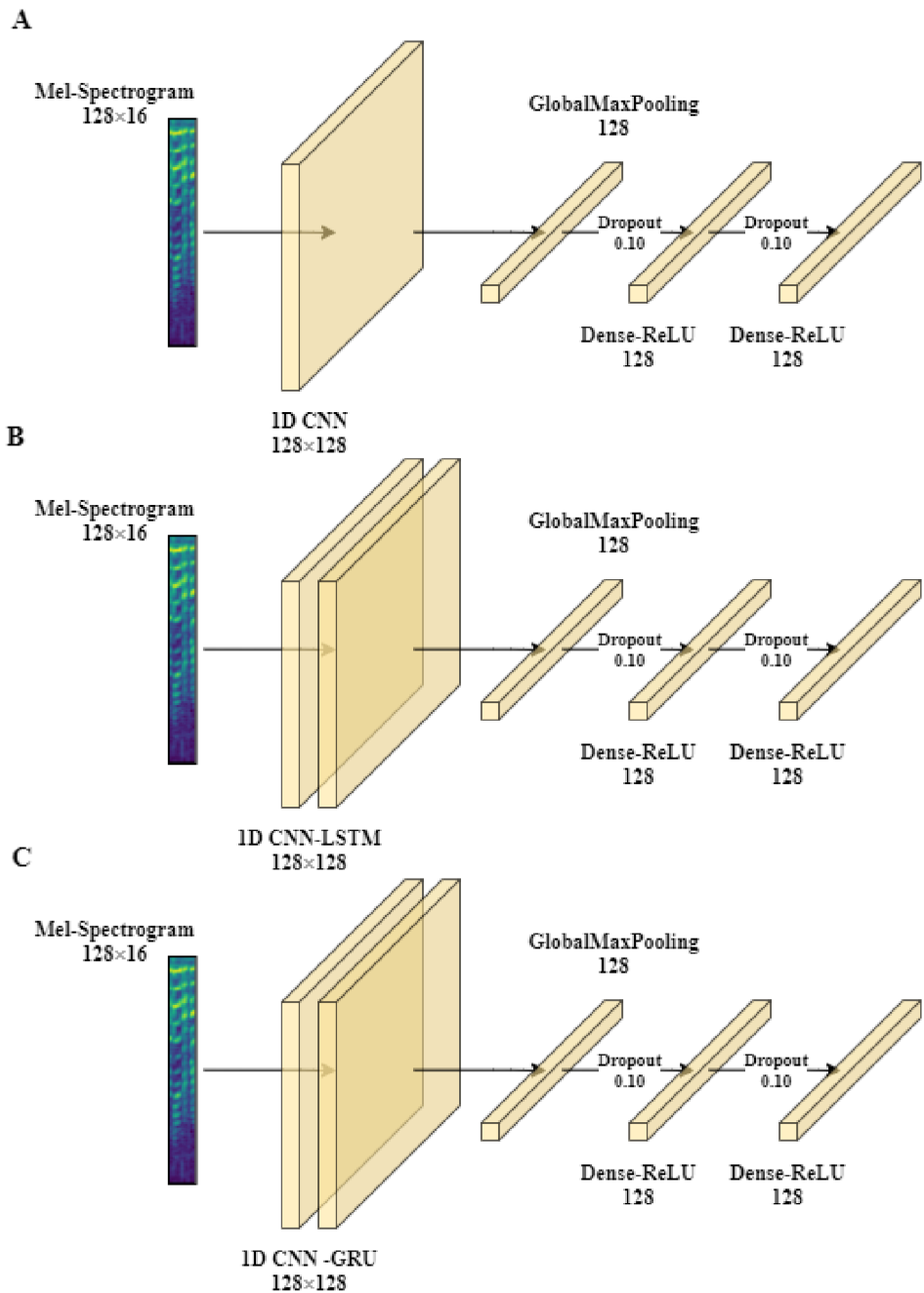

**Figure 3.** Architectures of the three encoders. (A) Encoder 1D CNN contains only 1 layer of 1D CNN that receives the 1-second long Mel-Spectrogram as an input. Meanwhile, for encoders (B) 1D CNN-LSTM and (C) 1D CNN-GRU, the input will be passed to 1D CNN then to LSTM and GRU, respectively. Next, in all encoders, the vocal data will be summarized by GlobalMaxPooling followed by dense layers with dropout rate of 0.10 which give out the output as feature vectors.

## 3.4. Hyperparameter Tuning

*Number of the Input Samples: 5, 10, 15, 30, 60 samples*

As previously stated, the experiment relies on the idea that probability for detecting markers associated with depression should increase when more random fragments are input into the model. However, we do not have information on the optimal number of this parameter. Thus, we plan to test the model with different numbers of voice samples.

*Size of the Kernels: 3, 5, 7*

A research conducted on depressed patients reveals that these groups of people have less change in vocal tract when speaking when compared with ordinary people [53][54][55]. Hence, they have weaker utterances (lower decibels than normal subjects), and lower emotional expression (frequency pattern that signals change in emotion) in their voice [55]. Regarding these, we are determined to identify the nature of frequency deviation in subjects with clear signs of depression. As we aim to identify these patterns, we do not know the area on the feature map that it covers. As a result, we decide to adjust the size of kernels to search for the configuration that suits the frequency pattern best.

As previously mentioned, we do not know that the frequency features that signal the disease are local or global patterns. Apart from that, we also want to further investigate and compare the importance of both local and global patterns in depression diagnosis. We have 3 types of encoders that we are planning to test our proposal:

*Type of Encoder: Encoder 1D CNN, Encoder 1D CNN-LSTM, Encoder 1D CNN-GRU.*

Every encoder has 1D CNN as its first layer for feature extraction. We want to use a 1D CNN-based encoder to determine the importance of local patterns in depressed and non-depressed subjects. On the other hand, for encoder 1D CNN-LSTM/GRU, we employ an additional RNN layer to analyze global patterns of the voice samples. We hypothesize that, apart from local features, global features may also play an important role in detecting the disease. We plan to not use only pure RNN encoder because we believe that RNN will perform better when receiving features from 1D CNN. For the process, the 1D CNN will first summarize association amongst acoustic features through local patterns; then, the RNN will examine the global pattern of the summarized acoustic features again. There are research papers indicating the ability of LSTM and GRU when handling short-time sequences [56][57]. So, we decide to compare the ability of these 2 RNNs by doing hyperparameter tuning to search for the one that performs the best.

As we want to cover all possible configurations, Standard Grid Search which is a standard training protocol is used in our project. Standard Grid Search emcompasses listing every combination of the three hyperparameters and applies them to the model. As a result, the model will be trained under all the parameters, and it will give out the scores representing each configuration performance. Hence, we can set the model with the parameters that have the highest classification power.

## 4. RESULTS

### 4.1. Training Strategy

We use 3-fold cross-validation to prevent bias in spliting dataset. To begin with, we divide our data into three subsets of similar sizes as depicted in Table 3. Every fold contains both depressed and non-depressed speakers (labeled with 1 and 0, respectively). In each round, 1 fold is selected for testing while the remaining is used for training.

Table 3. 3-fold cross-validation to partition DAIC-WoZ dataset

| K | Label | Number of Subjects |
|---|---|---|
| 0 | 0 | 30 |
| 0 | 1 | 60 |
| 1 | 0 | 42 |
| 1 | 1 | 17 |
| 2 | 0 | 37 |
| 2 | 1 | 21 |

In addition, we want the model to learn biomarkers associated with depression, not the features of each speaker. Hence, we employ stratified sampling to derive input fragments from the same person but selected randomly.

### 4.2 Results of Hyperparameter Tuning

Our classification results are analyzed based on 4 classification assessment methods: PR-AUC, F1 score, precision and recall. However, while tuning hyperparameters, we will focus mainly on PR-AUC scores to identify configurations with optimal performance. Here, three PR-AUC scores are generated from each subset from the 3-fold cross-validation. Subsequently, we use mean and standard error to calculate the value and confidence interval around the final PR-AUC score.

For the result of hyperparameter tuning, the CNN-based model generates the mean of PR-AUC score between 49.61 and 69.06. Meanwhile, both hybrid systems have much higher mean PR-AUC scores of 69.05 to 78.75 for CNN-LSTM, and 73.72 to 79.65 for CNN-GRU respectively. Regarding the result, we can conclude that the system that analyzes both local and global patterns is superior to the pure CNN model that focuses only on local features. Moreover, the combination of GRU and CNN outperforms the combination of CNN and LSTM (The hyperparameter tuning results of every configuration are written in Table S1). On the other hand, we find out that other parameters apart from the type of encoder (Kernel size and Number of samples) have little contribution to the improved performance. Finally, we discover that a model composed of 1D CNN-GRU with kernel size of 5 and 15 samples, generates the highest PR-AUC score 79.65 ± 2.02 (see Table 4). Hence, it has the best parameters, at least for this dataset, and we will use it to construct the final model.

Table 4. Best 5 configurations from hyperparameter tuning

| Sample Size | Kernel Size | Encoder Type | PR-AUC |
|---|---|---|---|
| **15** | **5** | **1D CNN -GRU** | **79.65** |
| 60 | 5 | 1D CNN - GRU | 79.41 |
| 30 | 7 | 1D CNN - GRU | 79.15 |
| 60 | 5 | 1D CNN-LSTM | 78.75 |
| 30 | 5 | 1D CNN - GRU | 78.48 |

### 4.3 Training and Post-modeling Analysis

Based on results in section 4.2., we train the final model using parameters in the selected configuration. Then, we assess model performance by producing ROC and PR curves from each fold of the dataset (see Figure 4). As a result, our graph shape corresponds with the one from a decent diagnostic test as supported by Kumar et al. [65]. Furthermore, Although some graphs (from each data fold) have minor shape differences between themselves, they all generate comparatively high AUC scores which signal an adequate classifying power. In addition, we also calculate the F1-score from every coordinate on the precision-recall graph to look for the point that yields the highest F1 score.

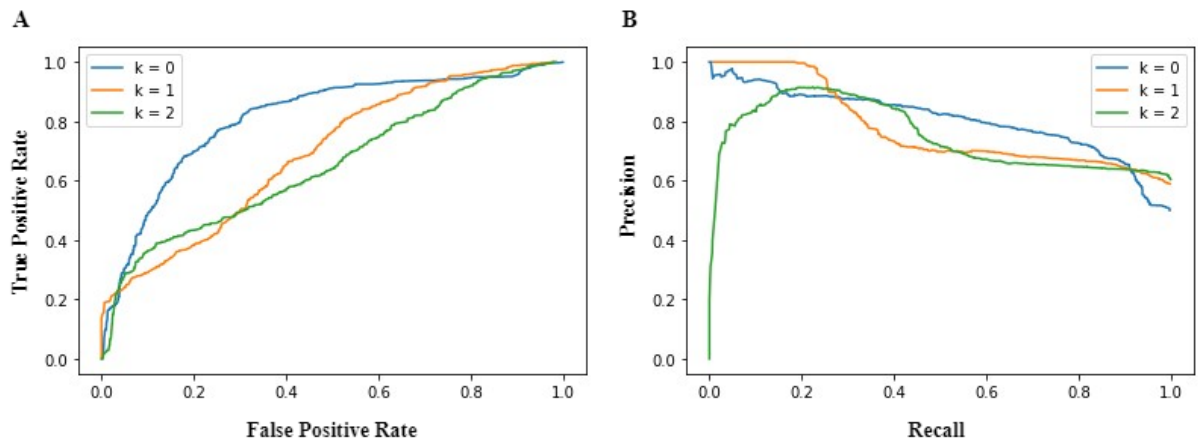

**Figure 4.** ROC and PR curve while training the final model (encoder 1D CNN-GRU with kernel size of 5 and 15 samples)

We also check whether the model has abilities to handle different levels of severity as depicted in Table 5 and in Figure S1. The result indicates that our model is capable of classifying depression from moderate to severe cases at relatively stable performance (i.e. F1 score 71.00-73.00; Precision 57.00-59.00; Recall 89.00-100.00) in all subclasses.

**Table 5.** Results of classification on different severity levels (based on PHQ-8 scores)

| PHQ - 8 Scores of Depressed sample | F1 Score | Precision | Recall |
|---|---|---|---|
| 10-14 | 71.00 | 57.00 | 92.00 |
| 15-19 | 71.00 | 59.00 | 89.00 |
| 20-24 | 73.00 | 57.00 | 100.00 |
| 10-24 | 75.00 | 64.00 | 91.67 |

After the final version of the model (1D CNN-GRU) is trained, we select 30,000 fragments with equal proportion of depressed and healthy subjects. Next, we extract the audio features from the fragments by using the 1D CNN-GRU encoder. Subsequently, the output features are plotted on the TSNE (T-Distributed Stochastic Neighbor Embedding) map where audio fragments with similar attributes are put together on the 2-dimensional plane. As depicted in Figure 5, there are generally three groups of sound fragments as clustered by K-means. If we colour samples by original labels (depressed: pink; healthy: blue), we find that 64.13% of fragments in Cluster A are from depressed subjects. Meanwhile, fragments in Cluster B and C contain fewer numbers of fragments from subjects with clear signs of depression (43.65% and 51.16%, respectively). Taking everything into account, findings in this section suggest that fragments in Cluster A which is an aggregate of audio samples from individuals with the disease might be an actual set of biomarkers useful for depression diagnosis.

**Figure 5. TSNE map of the extracted audio features -** From the 2D map, the coloured dots represent the audio feature extracted from each fragment. While blue represents fragments from healthy subjects, pink portrays fragments from depressed counterparts. (Colours appearing on these features are based on the original label of that person)

- In Cluster A: Depressed 64.13% and Healthy 35.87%
- In Cluster B: Depressed 43.65% and Healthy 56.35%
- In Cluster C: Depressed 51.16% and Healthy 48.84%

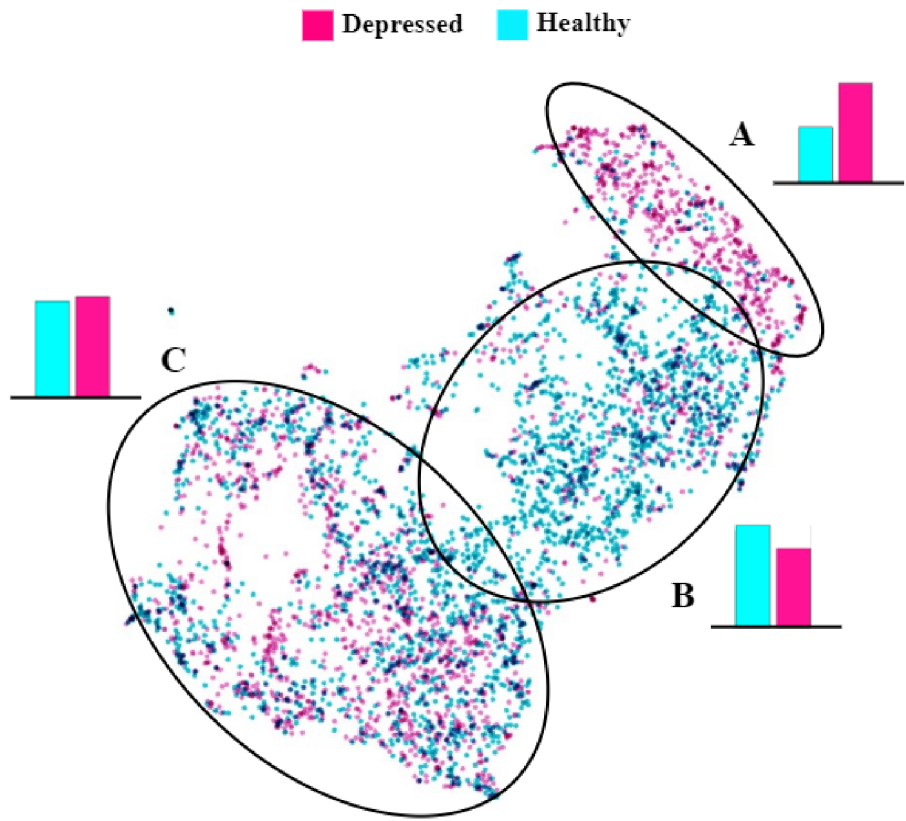

## 5. DISCUSSION

### 5.1. Comparison with Related Work

From the result, we discover that ability to classify depression patients corresponds significantly with the type of encoders whereas other parameters (kernel size and number of samples) have minor effects. Because the combination between CNN and RNN outperforms the pure CNN encoder, we hypothesize that global patterns are more necessary in detecting detection than local patterns. Other papers such as James R. Williamson et al. [53], and Jingying Wang et al. [54] propose that depressed and healthy subjects' voices have different acoustic patterns. The weak and low-pitch voice in this group of the population signals flatter frequency patterns in the acoustic plain when compared with healthy counterparts. Another piece of research that compares vocal attributes from both depressed and non-depressed subjects also supports this argument [55]. Therefore, we believe that there may be a correlation between the global patterns that we notice and the frequency patterns introduced by the aforementioned papers.

In addition, similarly to Peter T. Yamak et al. [56] and Shubham Khandelwal et al. [40], we found out that GRU exceeds LSTM when detecting patterns in a short time sequence.

Furthermore, if we compare our system with Esaú Villatoro-Tello et al. [58], their paper achieves 65.00 F1-Score whereas our model attains 75.00 F1 Score. Based on the same dataset, Esaú Villatoro-Tello et al. employs only CNN when searching for acoustic abnormality while our model is composed of CNN followed by GRU. Moreover, they use raw waveforms without converting it into spectrograms. Studying their procedure reassures that a hybrid CNN-RNN model is more likely to yield higher results than a pure CNN one. In addition, it also highlights our previously mentioned importance of global patterns over local patterns in the actual diagnosis.

If we take Adrián Vázquez-Romero et al. [43] into consideration, the major contrast is that they convert waveforms from the DAIC-WOZ dataset to the natural log of Short-time Fourier Transform (STFT) while we do not. Moreover, some of their hyperparameters are different, for instance, they tune the number of kernels in the CNN 1-D model while we fix it to 128 filters. For the result, this paper CNN-based prototype has the classification ability of 79.00 recall and 65.00 F1-score whereas our system outperforms them in both metrics. On the other hand, Afef Saidi et al. [59] constructs another kind of hybrid model that extracts depressed acoustic features using CNN and classifies such signals with SVM. Consequently, the model gains 69.00 F1 score, 65.00 precision, and 71.00 recall. In comparison, while Afef Said et al. classifies the vocal samples at a slightly better precision than us, our method surpasses their model in both F1 and recall (75.00 vs 69.00 and 91.67 vs 71.00 respectively) [59].

Table 5. F1-Score Comparison on Depressed Class

| Model | F1 Score | Precision | Recall |
|---|---|---|---|
| Esaú Villatoro-Tello et al. [57] | 65.00 | - | - |
| Megan V Smith et al. [14] | 68.69 | 62.00 | 77.00 |
| Adrián Vázquez-Romero et al. [43] | 65.00 | 55.00 | 79.00 |
| Afef Saidi et al. [59] | 69.00 | 65.00 | 71.00 |
| Cenk Demiroglu et al. [58] | 56.00 | 88.00 | 42.00 |
| Our Proposed Model | 75.00 | 64.00 | 91.67 |

Cenk Demiroglu et al.[58] applies the MRMR feature selection and SVR to search for depressed acoustic features in patients' audio waveform provided by DAIC-WOZ. The paper attains 56.00 F1 score and 42.00 recall which are less than our proposed model in both metrics (75.00 F1 score and 91.67 recall). Meanwhile, Cenk Demiroglu et al. generates 88.00 precision which exceeds 64.00 precision of our system. We hypothesize that they have a different approach to the disease diagnosis when compared with our project. While we underline our model's ability to screen depressed patients (addressed by higher recall), Cenk Demiroglu et al. emphasize precision above other aspects of the model. Nevertheless, PHQ-8 alone (Megan V Smith et al.) attains lower F1 score and recall than our system. [14]

### 5.2. 2D TSNE Feature Map Analysis

Based on our assumption, there are possibilities that the selected fragments contain no signal of depression, regardless of the sample's original status (depressed or healthy). For instance, samples selected from a person with severe depression may not contain indications of the disease. On the other hand, we also suspect that there are some acoustic patterns that are distinct to depressed people. Depiction in Figure 5 strengthens our hypothesis even further. In Figure 5, Cluster A contains fragments mostly from depressed subjects while the majority of fragments in Cluster B originate from healthy subjects. These clusters indicate those who have the disease and those who do not. This suggests that since features in cluster A are close to each other, they have similar characteristics that could signal depression (biomarkers). Such a discovery supports our earlier stated hypothesis about distinct frequency patterns.

Conversely, there are also some fragments from healthy subjects that have depression attributes, and some fragments from depressed subjects that have no biomarkers at all. Based on Figure 5, speech from those who have the disease may not always be enriched with biomarkers. On the contrary, the fragments that signal the symptom should spread sparsely along their recordings. Therefore, the aforementioned discovery supports our hypothesis that randomly selecting the vocal fragments from a subject will increase the chance of detecting abnormal frequency patterns.

In addition, there is another possibility that PHQ-8 does not correspond with a person's actual mental state. Concerning the fact that PHQ-8 is a screening questionnaire not a practical clinical diagnosis, there may be chances of encountering false positive samples. As a result, the presence of healthy subjects with depression characteristics in Cluster A might correspond with our aforementioned reasoning. Therefore, our model could be enhanced with information from approved clinical interviews and psychiatrists; so that our system will be capable of classifying false-labelled samples from the screening questionnaires in the future.

For Cluster B, although the majority of fragments in this group are healthy samples (56.35%), there is a slight difference between the proportion of both subjects. By analyzing the distribution of features inside Cluster B (in Figure 5), we hypothesize that fragments may contain sparse distribution of biomarkers. Moreover, if we take cluster C into consideration, there are also similar proportions of fragments from people with and without depression Based on the ratio of those who have the disease and ordinary subjects, we assume that Cluster C also shares little presence of vocal signatures that trace back to depression. The primary reason is our earlier hypothesis that biomarkers spread sparsely across the recording; thus, majority of the depressed subject's speech should be free of frequency abnormalities. Clear appearance of fragments from depressed people could be explained relying on earlier belief.

## 6. CONCLUSION & FUTURE WORK

In this paper, we have proposed the acoustic-based depression diagnostic system. While other papers address the prolonged frequency patterns across a period of time, we take the distinctive approach to find the indication of the disease. Our hypothesis relies on the fact that, during a small time interval, depressed subjects should possess abnormal frequency patterns which signal the ailment.

For the procedure, our model has analyzed both depressed and healthy samples from the DAIC-WOZ dataset by turning them into Mel Spectrograms. Additionally, random sampling is implemented to expand the opportunity of exposing unusual frequency. Moreover, we have tested our assumption with 3 types of encoders (1D CNN, 1D CNN-LSTM and 1D CNN-GRU), TimeDistributed Layer, and the fully connected layer with dropout regularizer. Apart from that, the model involves hyperparameter tuning which adjusts the prototype configuration until the optimum setting is found. Finally, three-fold cross validation is employed to augment the model training data and validating data.

The result has revealed that our system achieves an optimistic outcome in terms of AUC and F1-Score. It also outperforms the baseline procedure (PHQ-8) and some CNN-based depression detectors.

For future work, we are planning to utilize Layerwise Relevance Propagation (LRP) technique to indicate the features that contain depression vocal signatures [60]. Through reverse engineering the model will retrace the weight inside the dense layer until it reaches the destination with biomarkers. In this way, we will be able to help psychiatrists in their treatment of patients. We hope that our proposal will be a part to relieve this world from this mental disease, and no one should either suffer or lose their loved ones because of depression ever again.

## Acknowledgements

We would like to thank Mr. Thirawut Nilpanapan from Department of Mathematics and Computer, Faculty of Science, Chulalongkorn University, Ms. Saowaluk Changko from Biotec and Mr. Koravit Poysungnoen from Triam Udom Suksa School for proofreading our work. We also would like to thank The ICT Institute at University of California for generously sharing the DAIC-WoZ dataset with us.

Table S1. PR-AUC values of different model configurations.

| Sample Size | Kernel Size | Encoder Type | PR-AUC |
|---|---|---|---|
| 5 | 3 | 1D CNN | 55.82 |
| 5 | 3 | 1D CNN-LSTM | 69.05 |
| 5 | 3 | 1D CNN-GRU | 74.49 |
| 5 | 5 | 1D CNN | 56.99 |
| 5 | 5 | 1D CNN-LSTM | 74.88 |
| 5 | 5 | 1D CNN-GRU | 74.61 |
| 5 | 7 | 1D CNN | 56.01 |
| 5 | 7 | 1D CNN-LSTM | 73.40 |
| 5 | 7 | 1D CNN-GRU | 74.57 |
| 10 | 3 | 1D CNN | 56.86 |
| 10 | 3 | 1D CNN-LSTM | 72.74 |
| 10 | 3 | 1D CNN-GRU | 73.72 |
| 10 | 5 | 1D CNN | 56.18 |
| 10 | 5 | 1D CNN-LSTM | 76.46 |
| 10 | 5 | 1D CNN-GRU | 76.70 |
| 10 | 7 | 1D CNN | 55.96 |
| 10 | 7 | 1D CNN-LSTM | 74.59 |
| 10 | 7 | 1D CNN-GRU | 77.07 |
| 15 | 3 | 1D CNN | 55.97 |
| 15 | 3 | 1D CNN-LSTM | 77.23 |
| 15 | 3 | 1D CNN-GRU | 74.45 |
| 15 | 5 | 1D CNN | 59.32 |
| 15 | 5 | 1D CNN-LSTM | 74.80 |
| 15 | 5 | 1D CNN-GRU | 79.65 |
| 15 | 7 | 1D CNN | 59.01 |
| 15 | 7 | 1D CNN-LSTM | 76.12 |
| 15 | 7 | 1D CNN-GRU | 75.49 |

| Sample Size | Kernel Size | Encoder Type | PR-AUC |
|---|---|---|---|
| 30 | 3 | 1D CNN | 56.45 |
| 30 | 3 | 1D CNN-LSTM | 75.00 |
| 30 | 3 | 1D CNN-GRU | 76.58 |
| 30 | 5 | 1D CNN | 59.69 |
| 30 | 5 | 1D CNN-LSTM | 77.49 |
| 30 | 5 | 1D CNN-GRU | 78.48 |
| 30 | 7 | 1D CNN | 56.01 |
| 30 | 7 | 1D CNN-LSTM | 75.73 |
| 30 | 7 | 1D CNN-GRU | 79.15 |
| 60 | 3 | 1D CNN | 56.04 |
| 60 | 3 | 1D CNN-LSTM | 72.74 |
| 60 | 3 | 1D CNN-GRU | 75.63 |
| 60 | 5 | 1D CNN | 55.96 |
| 60 | 5 | 1D CNN-LSTM | 78.75 |
| 60 | 5 | 1D CNN-GRU | 79.41 |
| 60 | 7 | 1D CNN | 56.50 |
| 60 | 7 | 1D CNN-LSTM | 74.99 |
| 60 | 7 | 1D CNN-GRU | 77.71 |

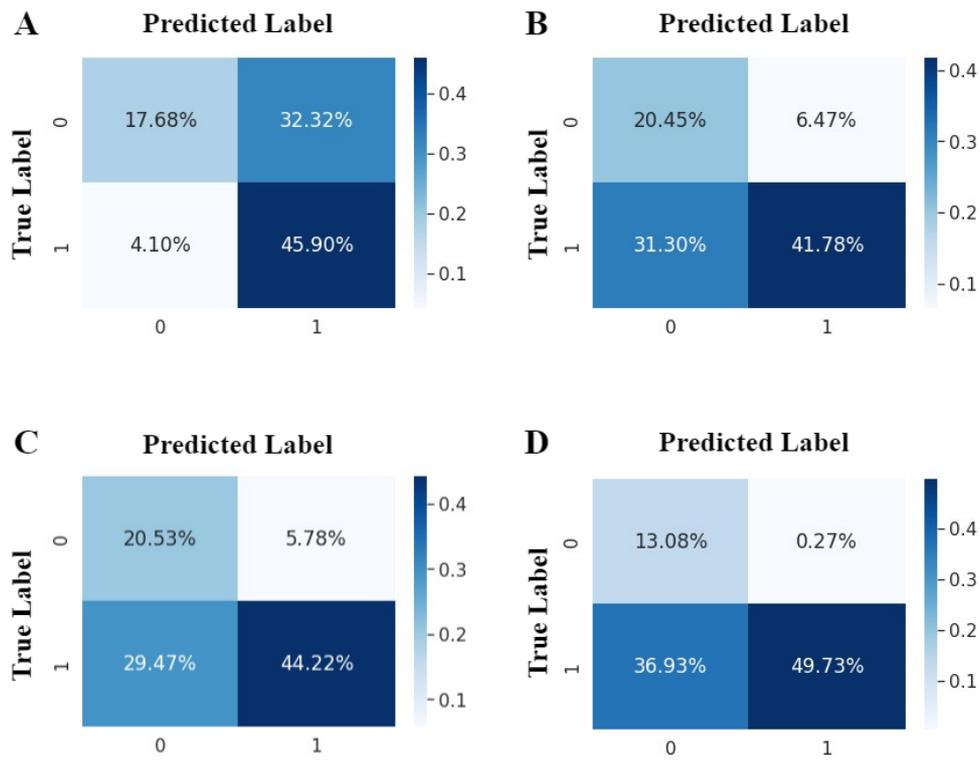

**Figure S1** Comparison of confusion matrices between true label and the predicted label. The matrices represent 4 different classes of depression, matrix A for PHQ-8 0-24 (all classes), B for PHQ-8 0-9 (minimal-mild) and 10-14 (moderate), C for PHQ-8 0-9 (minimal-mild) and 15-19 (moderately severe), and D for PHQ-8 0-9 (minimal-mild) and 20-24 (severe).